\documentclass{jetpl}
\usepackage{epsfig,cite}
\DeclareMathOperator{\Tr}{Tr} 
\twocolumn
\lat

\title{ Incoherent multiple Andreev reflection in an array of SNS junctions. }

\rtitle{Incoherent MAR in an array of SNS junctions. }

\author{N.\,M.~Chtchelkatchev$^{abc}$\thanks{e-mail: nms@itp.ac.ru},}
\rauthor{N.\,M.~Chtchelkatchev }

\sodauthor{N.\,M.~Chtchelkatchev}

\address{$^{a}$L.\,D.~Landau Institute for Theoretical Physics RAS,
117940 Moscow, Russia\\
$^{b}$ Institute for High Pressure Physics, Russian Academy of Sciences, Troitsk 142092, Moscow
Region, Russia \\
$^{c}$Moscow Institute of Physics and Technology,  Moscow 141700, Russia}
\dates{\today}*

\abstract{Last years many interesting effects related to incoherent MAR have been experimentally
found, but only few of them were theoretically explained. It was shown, for example, that if the
voltage at the edges of a linear array is $V$ then subgarmonic structures in the current -voltage
characteristics appear not only at usual for nonstationary Josephson effect positions,
$V_n=2\Delta/n$, where $n$ is integer, but also at voltages other than $V_n$.  A step towards
description of electron transport in a dirty array of SNS junctions is done in this letter. It is
shown that subgarmonic structures may indeed appear at ``unusual'' voltages.}

\PACS{73.23.-b, 74.45.+c, 74.81.Fa}

\begin{document}
\maketitle

Important role plays Andreev reflection mechanism in the subgap charge transfer through a normal
metal (N) -- superconductor (S) junction \cite{Andreev}. When an electron quasiparticle in a
normal metal with the energy below the superconduting gap reflects from the interface of the
superconductor into a hole, Cooper pair transfers into the superconductor. If the normal metal is
surrounded by superconductors, so we have a SNS junction, a number of Andreev reflections appear
at the NS interfaces. In equilibrium this leads to Andreev quasiparticle levels in the normal
metal that carry considerable part of the Josephson current; out of the equilibrium, when
superconductors are voltage biased, quasiparticles Andreev reflect about $2\Delta/eV$ times
transferring large quanta of charge ($\sim 2e\cdot [2\Delta/eV]$)  from one superconductor to the
other. This effect is called Multiple Andreev Reflection (MAR). If the voltage is near
$V_n=2\Delta/n$, where $n=1,2,\ldots$, so-called subgap features in current voltage
characteristics appear. Then large contribution to the current give quasiparticles that go from
the gap edge of one superconductor to the gap edge of the other superconductor (after MAR in the
normal region); bulk superconductor DoS is large at the gap and this is the reason of subgap
features.

This letter is devoted to investigation of electron transport in arrays of dirty superconductiong
mesoscopic SNS (SFS) junctions. I assume that normal parts of the junctions are ``long''. It
implies that minimum distance $d_0$ between adjacent superconductors is much larger than the
characteristic scale of anomalous green function (Cooper pair wave function) decay $\xi_N$ from a
superconductor in the normal metal. If the diffusion coefficient of the normal metal is $D_N$ than
$\xi_N\sim \sqrt{D_N/T}\ll d_0$, where $T$ is the temperature. Energy relaxation is not included
in the calculations. So it is also assumed that the array is not large, its characteristic length
does not exceed the characteristic length scale of quasiparticle energy relaxation in the normal
metal and in superconductors. [Quasiparticle energy relaxation was ``taken into account'' in
Ref.\cite{Shumeiko}, where incoherent MAR in long SNS junction was discussed, by small imaginary
part supplied to the energy in retarded and advanced greens functions; but collisional integrals
were not taken into account in kinetic equations. It is not clear why this procedure is correct. I
do not follow this way here.] Conditions listed above also mean that in equillibrium the proximity
(Josephson) effect between the superconducors is suppressed. Josephson current, for example, is
exponentially small with $d_0/\xi_N$. It is known that out of the equilibrium when there is a
finite bias between the superconducors the proximity effect restores in some sense: subgap
features appear due to MAR in current-voltage characteristics. MAR in long Josephson junctions
usually is referred to as `incoherent'' since there is no contribution to electron transport from
effects related to interference of quasiparticle wave functions in normal metals (contrary to
Josephson effect in ``short'' superconducting junctions). Last years many interesting effects
related to incoherent MAR have been experimentally found, but only few of them were theoretically
explained. It was shown, for example, that if the voltage at the edges of a linear array is $V$
then subgarmonic structures in the current -voltage characteristics appear not only at usual for
nonstationary Josephson effect positions, $V_n=2\Delta/n$, where $n$ is integer, but also at
voltages other than $V_n$, see Ref.\cite{Baturina} and refs. therein.  A step towards description
of electron transport in a dirty array of SNS junctions is done in this letter. It is shown that
subgarmonic structures may indeed appear at ``unusual'' voltages.
\begin{figure}[t]
\begin{center}
\includegraphics[height=55mm]{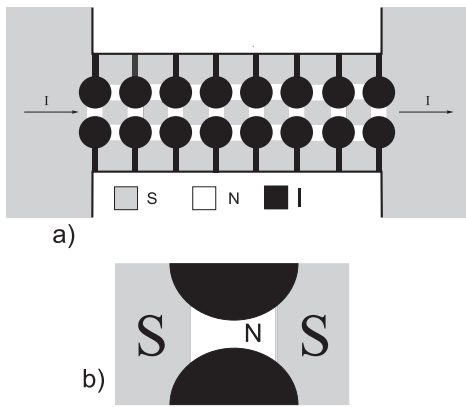}
\caption{Fig.1. a) An array of SNS junctions, like in experiments \cite{Baturina}. b) A sketch of
the normal layer connection to the superconductors in experimental SNS arrays. Black area is
insulating.} \label{fig:fig1}
\end{center}
\end{figure}
Investigation of electron transport is based on Usadel equations:
\begin{gather}\label{eq:Usadel_general}
    [\check{H},\check{G}]=iD\nabla\cdot \mathbf{\check{J}},\qquad
    \mathbf{\check{J}}=\check{G}\nabla \check{G},\qquad \check{G}^2=\check 1,
    \\
    \check{H}=\check 1(i\hat\sigma_z\partial_t-e\phi+\hat{\Delta}),
    \\
    \hat{\Delta}=
    \begin{pmatrix}
      0 &\Delta  \\
      -\Delta^*  & 0 \\
    \end{pmatrix},
    \\
    I(t,x)=\frac {\pi\hbar\sigma_{_N}}{4e}\Tr \hat\sigma_z \hat{J}^K(t,t;x),
\end{gather}
where
\begin{gather}
\check{G}=\begin{pmatrix}
      \hat{R} & \hat{K} \\
      0 & \hat{A} \\
    \end{pmatrix}.
\end{gather}
Here $\hat{R}$, $\hat{K}$ and $ \hat{A}$ denote retarded, Keldysh and advanced quasiclassical
green functions. The hat reminds that greens function are in turn matrices in Nambu space.
$\hat\sigma$ are Pauli matrices that act in Nambu space; $D$ is a diffusion coefficient that
equals $D_N$ in a normal metal and $D_S$ in a superconductor; $\Delta$ is the superconducting gap;
$\phi$ is electrical potential; $\sigma_N$ is the conductivity of a normal metal. The boundary
conditions for the Usadel equations at NS interfaces are:
\begin{gather}\label{eq:Boundary_cond_ap}
\sigma_{_{N_2}}\check {\mathbf {J}}_2\cdot \mathbf
{n}_2=\frac{G_{12}}{2}[\check{G}_2,\check{G}_1]_-,
\end{gather}
where $1,2$ label interface sides, $G_{12}$ is the surface conductance and $\mathbf n_2$ is the
unit normal to the interface pointing to the second half-space.

The problem is to calculate the current in a long SNS array using
Eqs.\eqref{eq:Usadel_general}-\eqref{eq:Boundary_cond_ap}.  Solving these equations directly is a
hard task because they are nonlinear, nonuniform and essentially time dependent (relative phases
of superconductors rotate with biases).  So the main task is developing an approach for the
problem in hand that allows to perform  significant part of transport calculations analytically
and that is applicable in rather wide range of system parameters. There is no universal approach
that helps to solve Usadel equations analytically, so any analytical method of Usadel equations
solution is usually  specific to the given class of the physical systems.

Normal layers in experimental SNS arrays \cite{Baturina} connect with superconductors like it is
shown in Fig.\ref{fig:fig1}. SNS junctions of this type are usually referred to as ``weak-links''
\cite{Likharev}. Boundary conditions, Eq.\eqref{eq:Boundary_cond_ap}, can be simplified in this
case: retarded and advanced Greens functions at superconducting sides of NS boundaries can be
substituted by Greens function from the bulk of the superconductors. These ``rigid'' boundary
conditions approximation are reasonable because a) the magnitude of the current is much smaller
than the critical current of the superconductor (this is assumed), b) the current entering the
superconductor from narrow normal metal wire [the width $\lesssim \xi=\sqrt{D_S/T_c}$, where $T_c$
is the critical temperature of the superconductor] spreads nearly at the NS interface over the
whole superconductor. There are  also other cases when rigid boundary conditions are correct, for
example, if the NS boundary has small transparency due to an insulator layer or other reasons.

The Keldysh greens function has the following general parametrization: $\hat{K}=\hat{R}\circ
\hat{f}-\hat{f}\circ \hat{A}$, where $\hat f$ is the distribution function. It was shown in
Ref.\cite{Shumeiko} that the current in a long SNS junction can be found from investigation of the
current distribution in an effective network where the role of voltages play distribution
functions made from components of $\hat f$, the role of resistances play NS resistances
renormalized by proximity effect and normal layer resistances. This idea can be applied to an SNS
array.

\begin{figure}[t]
\begin{center}
\includegraphics[height=25mm]{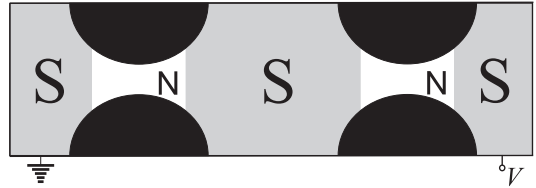}
\caption{Fig.2. The simplest array of SNS junctions: SNSNS. } \label{fig:SNSNS_scetch}
\end{center}
\end{figure}

It is convenient to write $\hat f=\hat 1 f_++\hat\sigma_z f_-$. Phases of the superconducting
order parameters rotate with the biases so one could write in general $\hat
f(\epsilon,\epsilon')=\sum_{n=\{n_i;i=1,2\ldots\}}\hat
f_n(\epsilon-\epsilon')\delta(\epsilon-\epsilon'+n_iV_i)$, where $n_i$ is integer, $V_i$ is the
bias at the $i$'s superconductor. (Similar consideration apply for retarded and advanced greens
functions.) However due to the absence of Josephson-type interference effects nonzero are only
that $\hat f_n$ which provide time independent (dissipative) component of the current as it is
well explained in Ref.\cite{Shumeiko}... One can write boundary conditions
Eq.\eqref{eq:Boundary_cond_ap} at an NS interface for $\hat f_n$ as
\begin{gather}\label{eq:boundary_f_pm_for convenience}
\sigma_N D_+\nabla f_+^{(2)}=-G_+(f_+^{(2)}-f_{+}^{(1)}),
\\
\sigma_N D_-\nabla f_-^{(2)}=-G_-(f_-^{(2)}-f_-^{(1)}).
\end{gather}
Here the  label 2 corresponds to the normal metal. For example,
\begin{gather}
D_+=\frac 1 4\Tr\left(\hat 1-\hat R\hat A\right),
\\
D_-=\frac 1 4\Tr\left(\hat 1-\hat R\hat\sigma_z\hat A\hat\sigma_z\right).
\end{gather}
Definitions of $G_\pm$ are similar and can be found, e.g., in Ref.\cite{Shumeiko}.

It is useful to go from $f_\pm$ to $n_{e(h)}$ \cite{Shumeiko} that are related as follows:
$f_+=1-(n_e+n_h)$, $f_-=(n_h-n_e)$. Then the boundary conditions, Eqs.\eqref{eq:boundary_f_pm_for
convenience}, can be written as
\begin{gather} \label{eq:Boundary_cond_exact_Ie}
\split I_e=G_T( n_e^{(1)}-&n_e^{(2)})-
\\
-G_A( &(n_e^{(2)}- n_h^{(2)}) -(n_e^{(1)}-n_h^{(1)})),\endsplit
\\ \label{eq:Boundary_cond_exact_Ih}
\split I_h=G_T( n_h^{(1)}-&n_h^{(2)})+
\\
+G_A(& (n_e^{(2)}-n_h^{(2)})-(n_e^{(1)}-n_h^{(1)})).\endsplit
\end{gather}
Here $G_T=G_+$, $G_A=(G_--G_+)/2$ and $I_{e(h)}=-\sigma_N\frac 1 2(D_+
\partial (n_e+ n_h)\pm D_- \partial (n_e- n_h))$. If the superconducting bank labelled by index ``1''
is in equilibrium, so $n_e^{(1)}=n_h^{(1)}=n_F$, then we arrive at boundary conditions written in
Eq.(21) of Ref.\cite{Shumeiko}. If one writes
Eqs.\eqref{eq:Boundary_cond_exact_Ie}-\eqref{eq:Boundary_cond_exact_Ih} for NS and SN interfaces
of a superconducting island and uses conditions of electron and heat currents conservation in the
island then quasiparticle distribution functions corresponding to superconducting islands can be
excluded from boundary conditions:
\begin{gather}\label{eq:I_e_boundary}
I_e=A_+(n_e^{(2)}-n_e^{(1)})+A_-(n_h^{(2)}-n_h^{(1)}),
\\\label{eq:I_h_boundary}
I_h=A_+(n_h^{(2)}-n_h^{(1)})+A_-(n_e^{(2)}-n_e^{(1)}).
\end{gather}
It is worth noting that Eqs.\eqref{eq:I_e_boundary}-\eqref{eq:I_h_boundary} are derived for a
linear array of SNS junctions; generalization of these equations is straightforward for more
complicated arrays.
\begin{figure}[th]
\begin{center}
\includegraphics[height=100mm]{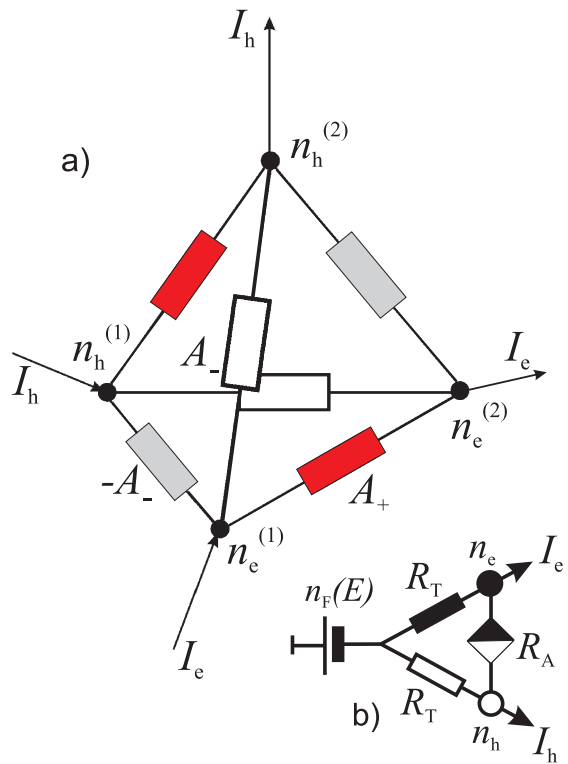}
\caption{Fig.3. a) An illustration of the boundary conditions
Eqs.\eqref{eq:I_e_boundary}-\eqref{eq:Boundary_cond_exact_Ih} in terms of a circuit is given in
Fig.\ref{fig_NSN_boundary_condition}. Electron and hole currents entering the left side of the
pyramid flow in one normal layer, the right currents flow in the other normal layer. $-A_-$
resistance describes Andreev reflection, $A_+$ --- quasiparticle normal transmission through the
superconductor and $A_-$ is Andreev transmission characteristics. b) An illustration of the
boundary conditions Eqs.\eqref{eq:I_e_boundary}-\eqref{eq:Boundary_cond_exact_Ih} (developed in
Ref.\cite{Shumeiko}) at the surface of the superconductor connected to an electron reservoir ,
i.e., when $n_e^{(1)}=n_h^{(1)}=n_F$ in the superconductor. Here $R_A=1/G_A$, $R_T=1/G_T$.}
\label{fig_NSN_boundary_condition}
\end{center}
\end{figure}
In the last pair of equations labels 1 and 2 correspond to normal layers surrounding the
superconducting island. Information that the island is superconducting is included in $A_\pm$
definition through $G_\pm$:
\begin{gather}
    A_\pm=\frac 1 2(\mu_+\pm\mu_-),
    \\
    \mu_\pm=\frac{G_\pm^{(1)}G_\pm^{(2)}}{G_\pm^{(1)}+G_\pm^{(2)}}.
\end{gather}
A circuit illustration of this boundary condition is given in
Fig.\ref{fig_NSN_boundary_condition}a. Electron and hole currents entering the left side of the
pyramid flow in one normal layer, the right currents flow in the other normal layer. $-A_-$
resistance describes Andreev reflection, $A_+$ --- quasiparticle normal transmission through the
superconductor and $A_-$ is Andreev transmission characteristics. It was assumed deriving
Eqs.\eqref{eq:I_e_boundary}-\eqref{eq:Boundary_cond_exact_Ih} that electrochemical potential
within the superconductor is constant. This is correct if the characteristic distance between NS
interfaces of the superconducting island is smaller than the electron disbalance characteristic
length $\lambda_Q$\cite{Tinkham}. It is implied that nonequillibrium quasiparticles above the gap
do not have enough time for energy relaxation at their fly through the superconductor, so
electrochemical potential within the superconductor is constant. Or if characteristic bias value
$\delta V$ between adjacent superconducting islands is much smaller than the gap and the
temperature is much below $T_c$ then most part of the current carry Cooper pairs through
superconductors rather than quasiparticles above the gap. Then correction to the total current in
the array from quasiparticle energy relaxation in superconductors is expected to be small as
$\delta V/\Delta\ll 1$ and discussion related to $\lambda_Q$ is not relevant.

Next important step is expressing all boundary conditions in terms of the distribution functions
$\bar n_{e(h)}$ \cite{Shumeiko}, where
\begin{gather}\notag
\overline{n}_\pm=n_\pm-\frac{I_\pm}{\sigma_N}\int_{0}^{\infty} dx'\left(\frac 1
{D_\pm(x')}-1\right)\equiv n_\pm-m_\pm I_\pm.
\end{gather}
Here $n_\pm=n_e\pm n_h$ and $\bar n_\pm=\bar n_e\pm \bar n_h$, $D_\pm \partial f_\pm\equiv
I_\pm/\sigma_N$. The integral here goes from the NS boundary ($x=0$) into the depth of the normal
metal. Boundary conditions Eqs.\eqref{eq:I_e_boundary}-\eqref{eq:Boundary_cond_exact_Ih} written
in terms of $\bar n_{e(h)}$ will have the same form if one replaces $G_\pm$ by $\bar
G_\pm={G_\pm}/({1+G_\pm m_\pm})$. Then
\begin{gather}
\bar\mu_\pm=\frac{\mu_\pm}{1+(m_\pm^{(2)}-m_\pm^{(1)})\mu_\pm},
\end{gather}
where indices 1,2 correspond to
two normal layers contacting with the superconducting island. It is useful to work with $\bar
n_{e(h)}$ because then the proximity effect renormalization of the boundary resistances, $m_\pm$,
is automatically taken into account.

The next step toward current calculation is to draw an effective network that describes MAR in the
array of the junctions using Eqs.\eqref{eq:I_e_boundary}-\eqref{eq:Boundary_cond_exact_Ih} and
evaluate partial currents in this network using to Kirchhoff's laws. Important task is to find
voltages at the superconducting islands. But this can be found easily in several cases. For
example, when the array consists of equal SNS junctions, or if most of the voltage drops at normal
layers. Below these situations will be discussed. More complicated cases I leave for extended
paper. If currents in the MAR network are found then electric current can be evaluated, e.g., as
follows:
\begin{gather}
I= \frac {1}{2e}\int dE \left(I_e(E)-I_h(E)\right),
\end{gather}
where  $I_{e(h)}$ correspond to the normal layer at the edge of the array contacting with the
superconducting reservoir with zero voltage.

\begin{figure}[t]
\begin{center}
\includegraphics[height=39mm]{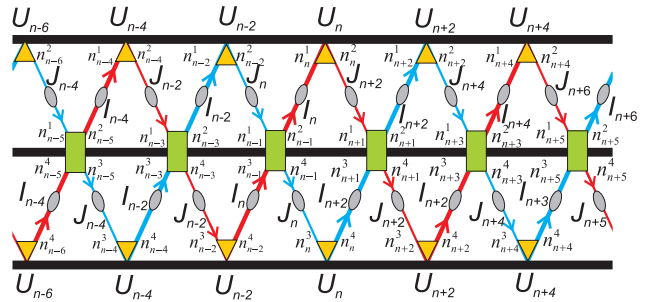}
\caption{Fig.4. MAR in a SNSNS array with equal SNS junctions. The graph shows the effective
circuit for quasiparticle currents in energy space. The role of voltages here play quasiparticle
distribution functions. For example $n_{n}^4$ is the quasiparticle distribution function depending
from $E+nV/2$; $U_n=n_F(E+nV/2)$. Boxes, triangles and ovals play the role of effective
resistances that come from Usadel equations and their boundary conditions,
Eqs.\eqref{eq:Usadel_general}-\eqref{eq:Boundary_cond_ap}. The oval is the resistance of the
normal layer. The box and the triangle correspond to the pyramidal bridge in
Fig.\ref{fig_NSN_boundary_condition}a and the three-terminal device in
Fig.\ref{fig_NSN_boundary_condition}b. The upper raw of $U$'s correspond to the first
superconductor of the array, the lower raw -- to the last superconductor. The thick line in the
center of the graph represents the superconducting island. } \label{fig:SNSNS}
\end{center}
\end{figure}
The format of the letter does not allow to describe complicated arrays here, so one of the
simplest SNS arrays, the SNSNS junction (Fig.\ref{fig:SNSNS_scetch}),  will be considered below as
an example. It will be shown how to construct an effective network that helps to describe its
transport properties. Transport in more complicated arrays, like in Fig.\ref{fig:fig1} can be
described in a similar manner as for SNSNS; it will be demonstrated in an extended version of this
paper.

Effective MAR network for a SNSNS junction like in  Fig.\ref{fig:SNSNS_scetch} is shown in
Fig.\ref{fig:SNSNS}. The bias between the supercondcutors at the edges of the array is $V$. Then
the bias of the central superconductor is $V/2$; it follows from symmetry reasons. Currents $I_n$
and $J_n$ correspond $I_e$ for lines beginning from $U_{n_1}$ and ending at $U_{n_2}$ with
$n_2>n_1$ and $-I_h$ vice-versa. The role of voltages play quasiparticle distribution functions.
For example $n_{n}^4$ is the quasiparticle distribution function depending from $E+nV/2$;
$U_n=n_F(E+nV/2)$. Bars above the distribution functions are supposed but not shown explicitly in
the figure. Boxes, triangles and ovals play the role of effective resistances (also with bars)
that come from Usadel equations and their boundary conditions,
Eqs.\eqref{eq:Usadel_general}-\eqref{eq:Boundary_cond_ap}. The oval is the resistance of the
normal layer. The box and the triangle correspond to the pyramidal bridge in
Fig.\ref{fig_NSN_boundary_condition}a and the three-terminal device in
Fig.\ref{fig_NSN_boundary_condition}b. The upper raw of $U$'s correspond to the first
superconductor of the array, the lower raw -- to the last superconductor. The thick line in the
center of the graph represents the superconducting island.

\begin{figure}[t]
\begin{center}
\includegraphics[height=46mm]{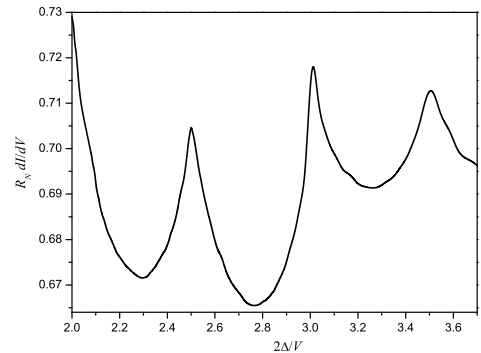}
\caption{Fig.5. Subharmonic structure in differential conductivity of SNSNS junction with the
ratio $r$ of SN boundary resistance to normal layer resistance $R_N$, $r=0.2$. Peaks at
half-integer $2\Delta/V$ do not appear in $dI/dV$ of SNS junctions.} \label{fig:graph}
\end{center}
\end{figure}

Lets find recurrence relations for the currents $I_n$ and $J_n$. It follows from
Fig.\ref{fig:SNSNS} that
\begin{gather}\notag
\begin{array}{rcl}
I_n &=& \bar G_T^{n-2}(n_{n-2}^4-U_{n-2})+\bar G_A^{n-2}(n_{n-2}^4-n_{n-2}^3),
\\
I_n&=&G_N(n_{n-1}^3-n_{n-2}^4),
\\
I_n&=&A_+^{n-1}(n_{n-1}^2-n_{n-1}^3)+A_-^{n-1}(n^1_{n-1}-n^4_{n-1}),
\\
I_n&=&G_N(n_n^1-n_{n-1}^2),
\\
I_n&=&\bar G_T^n(U_n-n_n^1)+\bar G_A^n(n_n^2-n_n^1),
\\
J_{n-2}&=&\bar G_A^{n-2}(n^4_{n-2}-n^3_{n-2})+\bar G_T^{n-2}(U_{n-2}-n^3_{n-2}),
\\
J_{n+2}&=&\bar G_A^{n}(n^2_{n}-n^1_{n})+\bar G_T^{n}(n_{n}^2-U_n);
\end{array}
\\\notag
\begin{array}{rcl}
J_n &=& \bar G_T^{n-2}(n_{n-2}^2-U_{n-2})+\bar G_A^{n-2}(n_{n-2}^2-n_{n-2}^1),
\\
J_n&=&G_N(n_{n-1}^1-n_{n-2}^2),
\\
J_n&=&A_+^{n-1}(n_{n-1}^4-n_{n-1}^1)+A_-^{n-1}(n^3_{n-1}-n^2_{n-1}),
\\
J_n&=&G_N(n_n^3-n_{n-1}^4),
\\
J_n&=&\bar G_T^n(U_n-n_n^3)+\bar G_A^n(n_n^4-n_n^3),
\\
I_{n-2}&=&\bar G_A^{n-2}(n^2_{n-2}-n^1_{n-2})+\bar G_T^{n-2}(U_{n-2}-n^1_{n-2}),
\\
I_{n+2}&=&\bar G_A^{n}(n^4_{n}-n^3_{n})+\bar G_T^{n}(n_{n}^4-U_n).
\end{array}
\end{gather}
Getting rid of the distribution functions in these set of equations one can get:
\begin{multline}\label{eq:I_recurrent}
I_n\left(\frac 1{A^+_{n-1}-A_{n-1}^-}+(a_n+a_{n-2})\right)-
\\
-I_{n-2}b_{n-2}- I_{n+2}b_n =(U_n-U_{n-2}).
\end{multline}
Same equation satisfies $J_n$. Here
\begin{gather}
a_n=\frac 1 {G_N}+\frac 1 {\bar G_T^{n}}-\frac{1}{2\bar G_A^{n}+\bar G_T^{n}}\frac{\bar
G_A^{n}}{\bar G_T^{n}},
\\
b_n=\frac{1}{2\bar G_A^{n}+\bar G_T^{n}}\frac{\bar G_A^{n}}{\bar G_T^{n}}.
\end{gather}
Eq.\eqref{eq:I_recurrent} coincides with recurrence relation Eq.(52) from Ref.\cite{Shumeiko}
derived for a single SNS junction if I replace in Eq.(52) $G_N$ by
\begin{gather}\label{eq:G_N/G_-}
\widetilde{G_N}=\frac {G_N} 2\frac{{\bar G_-}}{G_N+{\bar G_-}}.
\end{gather}
It means that SNSNS array behaves similarly as a single SNS junction but with energy dependent
resistance of the normal layer. $\bar G_-$ has singularities at energy corresponding to the gap
edges of the superconducting island in the center of the SNSNS array. This is the reason of
subharmonic singularities in current voltage characteristics if $2\Delta/V=n/2$, $n=1,2,\ldots$
instead of ``usual'' positions,$2\Delta/V=n$; Fig.\ref{fig:graph} illustrates it.  From
Eq.\eqref{eq:G_N/G_-} follows that unusual position of subharmonic singularities disappear if
$\tilde G_N\to G_N$ if the resistance of the normal layer exceeds the resistance of SN interfaces.
Then the central superconducting island of the SNSNS array effectively ``disappear''. It was
checked if the exchange field in SFSFS junction splits subharmonic structure; there was found no
exchange field splitting effects because configurations of the ferromagnet and the superconductor
like in Fig.\ref{fig:fig1} does not lead to enough exchange field deformation of superconductor
DoS near SF boundary that is necessary for the splitting effect observation. This paragraph is
conclusion of the paper.

I'm grateful to T.Baturina and I.S. Burmistrov for stimulating discussions. I especially thank
I.~Drebushchak for helpful discussions and the idea to draw Fig.\ref{fig_NSN_boundary_condition}.
I also thank RFBR 03-02-16677, the Russian Ministry of Science, the Netherlands Organization for
Scientific Research (NWO), CRDF and Russian Science Support foundation.


\begin{thebibliography}{99}
\bibitem{Andreev} A.F.Andreev, Zh. \'{E}ksp. Teor. Fiz. \textbf{46}, 1823 (1964) [Sov. Phys. JETP
\textbf{19}, 1228 (1964)].

\bibitem{Baturina} T.I. Baturina, Yu. A. Tsaplin, A.E. Plotnikov \textit{et al}, JETP Lett.
\textbf{81}, 10 (2005); T.I. Baturina, D.R. Islamov, Z.D. Kvon, JETP Lett. \textbf{75}, 326
(2002). T.I. Baturina, Z.D. Kvon, A.E. Plotnikov, Phys. Rev. B \textbf{63}, 180503 (2001).
\bibitem{Likharev} K.K. Likharev, Rev. Mod. Phys. \textbf{51}, 101 (1979).

\bibitem{Shumeiko} E.V. Bezuglyi, E.N. Bratus', V.S. Shumeiko \textit{et al},
Phys.Rev. B \textbf{62}, 14439 (2002).

\bibitem{Tinkham} M.Tinkham, Introduction to superconductivity, Mc.Graw-Hill Inc., 1996.

\end{thebibliography}
\end{document}